\documentclass[preprint,12pt]{elsarticle}
\usepackage{braket}
\usepackage{graphicx}
\usepackage{amssymb}
\usepackage{amsmath}
\usepackage{dsfont}
\usepackage{listings}
\usepackage[normalem]{ulem}

\newcounter{bla}

\usepackage{xcolor}
\usepackage{upgreek}

\journal{Computer Physics Communications}

\begin{document}
	
	\begin{frontmatter}
		
		\title{Diatomic-py: A python module for calculating the rotational and hyperfine structure of $^{1}\Sigma$ molecules}
		
		\author{Jacob~A.~Blackmore\textsuperscript{1,2}\corref{jake}}
		\author{Philip~D.~Gregory\textsuperscript{1}}
		\author{Jeremy~M.~Hutson\textsuperscript{3}}
		\author{and~Simon~L.~Cornish\textsuperscript{1}\corref{Simon}}
		
		\cortext[jake]{ \textit{E-mail address:} jacob.blackmore@physics.ox.ac.uk}
		\cortext[Simon] {\textit{Corresponding author. E-mail address:} s.l.cornish@durham.ac.uk}
		\address{\textsuperscript{1} Department of Physics, Durham University, South Road, Durham, United Kingdom}
		\address{\textsuperscript{2} Department of Physics, University of Oxford, Parks Road, Oxford, United Kingdom}
		\address{\textsuperscript{3} Department of Chemistry, Durham University, South Road, Durham, United Kingdom}

\begin{abstract}
We present a computer program to calculate the quantised rotational and
hyperfine energy levels of $^{1}\Sigma $ diatomic molecules in the presence
of dc electric, dc magnetic, and off-resonant optical fields. Our program
is applicable to the bialkali molecules used in ongoing state-of-the-art
experiments with ultracold molecular gases. We include functions for the calculation
of space-fixed electric dipole moments, magnetic moments and transition
dipole moments.
\end{abstract}

\end{frontmatter}

{\bf PROGRAM SUMMARY}
\begin{small}
\noindent
{\em Program Title:Diatomic-Py}                                          \\
{\em CPC Library link to program files:} (to be added by Technical Editor) \\
{\em Developer's respository link: DOI: 10.5281/zenodo.6632148 }  \\
{\em Code Ocean capsule:} (to be added by Technical Editor)\\
{\em Licensing provisions:} BSD 3-clause\\
{\em Programming language: Python $\ge$ 3.7} \\
{\em Nature of problem: Calculation of the rotational and hyperfine structure of $^1\Sigma$ molecules in the presence of dc magnetic, dc electric, and off-resonant laser fields.}\\
{\em Solution method: A matrix representation of the Hamiltonian is constructed in the uncoupled basis set. Eigenstates and eigenenergies are calculated by numerical diagonalization of the Hamiltonian.} \\
\\
{\em Additional comments including restrictions and unusual features: Restricted to calculating the Stark and Zeeman shifts with co-axial electric and magnetic fields. }
\\
\end{small}

\section{Introduction}
\label{sec:intro}

A new generation of experiments is now able to produce ultracold $^{1}\Sigma $ ground-state molecules by associating pairs of alkali atoms~\cite{Deiglmayr2008,Ni2008,Aikawa2010,Danzl2010,Takekoshi2014,Molony2014,Park2015,Guo2016,Seesselberg2018,Hu2019,Yang2019,Voges2020,Cairncross2021,Lam2022}. Precise knowledge of the rotational and hyperfine structure is critical
for nearly all the foreseeable applications of these molecules (see e.g.~\cite{Carr2009}), and particularly for those that require long-lived quantum coherences between
states~\cite{Neyenhuis2012,Park2017,Blackmore2018,Caldwell2020,Burchesky2021,Gregory2021}. Basic calculation of the rotational structure of diatomic molecules is well understood~\cite{BrownandCarrington}, but the addition of hyperfine structure, together with external fields that prevent conservation of total angular momentum, greatly increases the size of the Hilbert space over which calculations must be performed. Experimental studies of ultracold molecules often involve external dc magnetic and electric fields and often also need an off-resonant optical field for trapping.
There is a need in the ultracold-molecule community for accessible
and open-source tools to perform these calculations.

In this work we present a flexible Python-based program for calculating the rotational and hyperfine structure of $^{1}\Sigma $ molecules in external electromagnetic fields. Our code contains a module to automate the construction of the Hamiltonian, which is then diagonalized using functions from the numpy stack~\cite{NumpyPaper,ScipyPaper}. We also include functions that simplify the calculation of important quantities such as the electric dipole moments for transitions between pairs of states. We give example plots that demonstrate use of the code to calculate Zeeman, dc Stark, and ac Stark maps for the hyperfine states of $^{1}\Sigma $ molecules.

\section{Theoretical background}
\label{sec:theory}

The model we consider is valid for diatomic molecules in a single vibrational state of a $^{1}\Sigma $ electronic state, as is relevant to most current experiments with ultracold bialkali molecules. For generality, we label the molecule AB, where A and B are the component nuclei. The Hamiltonian of such a molecule is \cite{Aldegunde2008}
\begin{equation}
H_{\mathrm{AB}} = H_{\mathrm{rot}} + H_{\mathrm{hf}} + H_{
\mathrm{ext}},
\label{eq:total_ham}
\end{equation}
where $H_{\mathrm{rot}}$ describes the rotational structure, $H_{\mathrm{hf}}$ describes the hyperfine structure and $H_{\mathrm{ext}}$ describes the interaction between the molecule and external fields.

We construct the Hamiltonian in the fully uncoupled basis where the relevant
angular momentum quantum numbers are the molecule's rotational angular
momentum ($N$) and the spin of the two nuclei ($i_{\mathrm{A}}$,
$i_{\mathrm{B}}$). Each of these angular momenta also has a projection
onto the $z$ axis of our coordinate system, defined by the direction of
the magnetic field, giving six quantum numbers in total: $N$,
$M_{N}$, $i_{\mathrm{A}}$, $m_{\mathrm{A}}$, $i_{\mathrm{B}}$,
$m_{\mathrm{B}}$. As we do not consider effects that could change the nuclear
spin (only its projection) where relevant we write our basis states as
$\ket{N,M_{N},m_{\mathrm{A}},m_{\mathrm{B}}}$ for brevity. We also define
the vector operators $\boldsymbol{N}$, $\boldsymbol{I}_{\mathrm{A}}$ and
$\boldsymbol{I}_{\mathrm{B}}$ that are associated with each of these angular
momenta.

In the following subsections, we give the mathematical expressions that describe each of the components of the Hamiltonian~\ref{eq:total_ham}.

\subsection{The rotational Hamiltonian, $H_{\mathrm{rot}}$}
\label{sec2.1}

We describe the rotation of the molecule using a rigid-rotor model; this results in a spectrum of rotational states that have energy $E_{N}\approx B_{v} N(N+1)$, where $B_{v}$ is the rotational constant of the molecule in vibrational state $v$. In terms of angular momentum operators, the rotational contribution to the Hamiltonian is~\cite{BrownandCarrington}
%
\begin{equation}
H_{\mathrm{rot}} = B_{v}( \boldsymbol{N}\cdot \boldsymbol{N}) - D_{v}(
\boldsymbol{N}\cdot \boldsymbol{N})^{2}.
\end{equation}
This includes a small correction for centrifugal distortion, characterised by the distortion coefficient $D_{v}$.

\subsection{The hyperfine Hamiltonian, $H_{\mathrm{hf}}$}
\label{sec2.2}

The hyperfine component of the Hamiltonian is necessary for molecules with non-zero nuclear spin. It describes interactions between the angular momentum of the two nuclei, and between each nucleus and the rotational angular momentum of the molecule. We can split this component further into terms that describe four different interactions,

\begin{equation}
H_{\mathrm{hf}} = H_{\mathrm{quad}}+H^{(0)}_{\mathrm{spin-spin}}+H^{(2)}_{
\mathrm{spin-spin}}+H_{\mathrm{spin-rotation}}.
\end{equation}
The first term $H_{\mathrm{quad}}$ describes the nuclear electric quadrupole interaction and is written as
\begin{equation}
H_{\mathrm{quad}} = \sum _{j=\mathrm{A},\mathrm{B}} e
\boldsymbol{\mathrm{{Q}}}_{j} \cdot \boldsymbol{\mathrm{{q}}}_{j},
\label{eq:NuclearQuad}
\end{equation}
where $e\boldsymbol{\mathrm{Q}}_{j}$ and $\boldsymbol{\mathrm{q}}_{j}$ are spherical tensors of rank 2 representing the nuclear quadrupole moment and electric field gradient (at the position of the nucleus) of nucleus $j$. This component of the Hamiltonian is governed by the molecular coupling constants $(eQq)_{\mathrm{A}}$ and $(eQq)_{\mathrm{B}}$ that contain both the magnitude of the nuclear electric quadrupole moments and the electric field gradients relevant for nuclei A and B, respectively; note this term is non-zero only when
$i_{j}>1/2$.

The second and third terms describe scalar and tensor interactions between the nuclear spins,

\begin{subequations}
\begin{equation}
H^{(0)}_{\mathrm{spin-spin}} =c_{4} \boldsymbol{I}_{\mathrm{A}}
\cdot \boldsymbol{I}_{\mathrm{B}},
\end{equation}
\begin{equation}
H^{(2)}_{\mathrm{spin-spin}}=-c_{3}\sqrt{6}\boldsymbol{\mathrm{T}}^{2}(C)
\cdot \boldsymbol{\mathrm{T}}^{2}\left (\boldsymbol{I}_{\mathrm{A}},
\boldsymbol{I}_{\mathrm{B}}\right ).
\label{eq:tensor_ss}
\end{equation}
\end{subequations}
This depends on the molecular constants $c_{3}$ and $c_{4}$, and a pair of second-rank tensors $\boldsymbol{\mathrm{T}}^{2}$ that describe the angular dependence of the interactions. In the above the tensor $\boldsymbol{\mathrm{T}^{2}}(C)$ has components given by ${T^{2}_{q}(C) =C^{2}_{q}(\theta ,\phi )}$, where $C^{2}_{q}$ is a Racah-normalised\footnote{$C^{k}_{0}(0,0)=1$.} spherical harmonic of order 2. The arguments of each component are the polar angle $\theta $ of the molecule's internuclear axis ($\boldsymbol{\hat{n}}$) from $z$ and the azimuthal angle $\phi $. The Racah-normalised forms of the spherical harmonics are related to the $\mathcal{L}^{2}$-normalised versions ($Y^{k}_{q}$) through ${C^{2}_{q}(\theta ,\phi ) = \sqrt{4\pi /5}Y^{2}_{q}(\theta ,\phi )}$.

The tensor $\boldsymbol{\mathrm{T}}^{2}( \boldsymbol{I}_{A} ,\boldsymbol{I}_{B} )$ is the second-rank tensor product of the two vectors $\boldsymbol{I}_{\mathrm{A}}$ and $\boldsymbol{I}_{\mathrm{B}}$. This can be written as

\begin{subequations}
\begin{gather}
{T}^{2}_{\pm 2}\left (\boldsymbol{I}_{\mathrm{A}},\boldsymbol{I}_{
\mathrm{B}}\right )= I^{\mathrm{A}}_{\pm 1}I^{\mathrm{B}}_{\pm 1},
\\
{T}^{2}_{\pm 1}\left (\boldsymbol{I}_{\mathrm{A}},\boldsymbol{I}_{
\mathrm{B}}\right )= \frac{1}{\sqrt{2}} \left ( I^{\mathrm{A}}_{\pm 1}
I^{\mathrm{B}}_{0} +I^{\mathrm{A}}_{0} I^{\mathrm{B}}_{\pm 1}\right ),
\\
{T}^{2}_{0}\left (\boldsymbol{I}_{\mathrm{A}},\boldsymbol{I}_{
\mathrm{B}}\right )=\frac{1}{\sqrt{6}} \left (I^{\mathrm{A}}_{+1} I^{
\mathrm{B}}_{-1} + I^{\mathrm{A}}_{-1} I^{\mathrm{B}}_{+1} +2 I^{
\mathrm{A}}_{0} I^{\mathrm{B}}_{0} \right ),
\\
\end{gather}
\end{subequations}
where $I^{j}_{q}$ are the components of $\boldsymbol{I}_{j}$.

The final term describes spin-rotation interactions that arise due to the magnetic moment of each nucleus interacting with the magnetic field generated by the rotating molecule. This term is given by
\begin{equation}
H_{\mathrm{spin-rotation}} = \sum _{j=\mathrm{A}, \mathrm{B}} c_{j}
\boldsymbol{N} \cdot \boldsymbol{I}_{j},
\end{equation}
where $c_{j}$ is the coupling constant for nucleus $j$.

\subsection{Interaction between the molecule and external fields, $H_{\mathrm{ext}}$}
\label{sec2.3}

To interpret current experiments with ultracold molecules, we need to calculate the internal structure in the presence of external electromagnetic fields. We further decompose the component of the Hamiltonian that describes the interactions between the molecule and external fields as
\begin{equation}
H_{\mathrm{ext}} = H_{\mathrm{Z}} + H_{\mathrm{dc}} + H_{\mathrm{ac}},
\end{equation}
where $H_{\mathrm{Z}}$, $H_{\mathrm{dc}}$, and $H_{\mathrm{ac}}$ describe the interaction with dc magnetic, dc electric, and non-resonant optical fields, respectively.

To describe the effect of a dc magnetic field $\boldsymbol{B}$, we construct the Hamiltonian
\begin{equation}
H_{\mathrm{Z}}=-g_{\mathrm{r}} \mu _{\mathrm{N}} \boldsymbol{N}
\cdot \boldsymbol{B}-\sum _{j=\mathrm{A},\mathrm{B}} g_{j}\left (1-
\sigma _{j}\right ) \mu _{\mathrm{N}} \boldsymbol{I}_{j} \cdot
\boldsymbol{B}.
\label{eq:Full_Zeeman}
\end{equation}
Here the first term accounts for the magnetic moment generated by the rotation of the molecule, described by the rotational $g$-factor $g_{\mathrm{r}}$. The second term accounts for the magnetic moments associated with the nuclear spins, characterised by the nuclear $g$-factors $g_{j}$, shielded isotropically by a factor~$\sigma _{j}$.

For polar molecules, dc electric fields couple strongly to the rotational angular momentum of the molecule. They can be used to orient the molecule in the laboratory frame, resulting in space-fixed dipoles that produce strong interactions over long range. The coupling between a dc electric field $\boldsymbol{E}_{\mathrm{dc}}$ and the angular momentum of the molecule is described by the Hamiltonian
\begin{equation}
H_{\mathrm{dc}}=-\mu _{0}\boldsymbol{E}_{\mathrm{dc}}\cdot
\hat{\boldsymbol{n}},
\label{eq:DC_Ham1}
\end{equation}
where $\mu _{0}$ is the magnitude of the electric dipole moment in the frame of the molecule and $\hat{\boldsymbol{n}}$ is a unit vector that points along the internuclear axis of the molecule.

Finally, interactions between the molecule and off-resonant optical fields are important for molecules confined to optical traps. Here, there is an interaction between the molecule and the oscillating electric field $\boldsymbol{E}_{\mathrm{ac}}$ of the trap light. For linearly polarised light the Hamiltonian is
\begin{equation}
H_{\mathrm{ac}}= -\frac{1}{2} \boldsymbol{E}_{\mathrm{ac}}\cdot
\boldsymbol{\upalpha}\cdot \boldsymbol{E}_{\mathrm{ac}},
\label{eq:AC_Ham}
\end{equation}
where $\boldsymbol{\upalpha}$ describes the molecular polarisability tensor, which depends on the wavelength of the light. The magnitude of $\boldsymbol{E}_\mathrm{ac}$ is related to the laser intensity by $I_\mathrm{ac} = |\boldsymbol{E}_\mathrm{ac}|^2/(\epsilon_0 c)$. The polarisability of the molecules is generally anisotropic; for a molecule oriented at an angle $\theta $ to the laser polarisation, defined by the angle $\beta $ from the $z$ axis, the polarisability along the axis of polarization is
\begin{equation}
\label{eq:Polarizability}
\begin{split} \alpha (\theta ) &= \alpha _{\parallel}\cos ^{2}\theta +
\alpha _{\perp}\sin ^{2}\theta
\\
&= \alpha ^{(0)} + \alpha ^{(2)}P_{2}(\cos \theta ),
\end{split}
\end{equation}
where $\alpha ^{(0)}$ and $\alpha ^{(2)}$ describe isotropic and anisotropic components of the molecular polarisability and $P_{2}(x) =(3x^{2}-1)/2$ is the second-order Legendre polynomial.

\section{Program structure}
\label{sec:structure}

The Diatomic-py package has two key modules: \texttt{hamiltonian} and \texttt{calculate}. The former contains functions used to construct the Hamiltonian, while the latter allows the efficient calculation of key quantities from the eigenvalues and eigenstates found by diagonalizing the Hamiltonian.

\subsection{Hamiltonian}
\label{sec3.1}

The Hamiltonian matrix is a two-dimensional, square array of dimension $(2\times i_{\mathrm{A}}+1)\times (2\times i_{\mathrm{B}}+1)\times (N_{\mathrm{max}}+1)^{2}$. To create this object, the user supplies a Python dictionary containing the relevant molecular constants and a value for $N_{\mathrm{max}}$, the highest-energy rotational state to include in the basis set, to the function \texttt{build\_hamiltonians}. We include current values of the constants for a selection of experimentally relevant bialkali molecules from~\cite{Neyenhuis2012,Gregory2016,Aldegunde2017}.

\begin{figure*}[t]
\includegraphics[width=\textwidth]{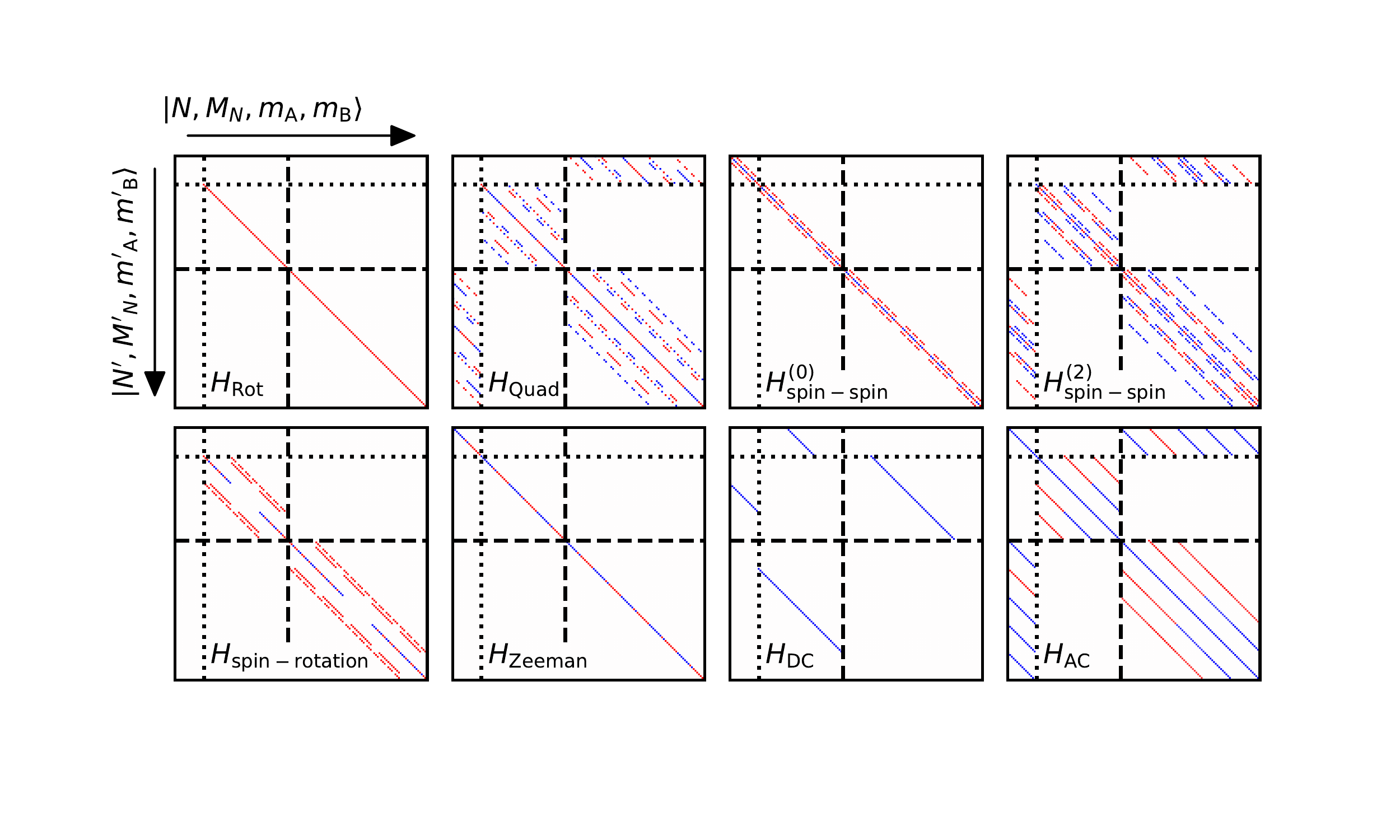}
\vspace{-2cm}
\caption{Contributing terms to the molecular Hamiltonian are shown graphically for rotational states $N=0,1,2$ and nuclear spins $i_{\mathrm{A}}=i_{\mathrm{B}}=3/2$. Each contribution consists of a 2D square matrix with $\sum _{N=0,1,2}(2N+1)(2i_{\mathrm{A}}+1)(2i_{\mathrm{B}}+1)=144$ elements in each direction. Non-zero matrix elements are indicated by the coloured markings, with positive values designated red and negative values blue. The dotted and dashed lines show the boundaries between $N=0,1$ and $N=1,2$ states respectively. The magnetic field $\boldsymbol{B}$ and the dc electric field $\boldsymbol{E}_{\mathrm{ac}}$ are coaxial and point along the $z$ axis. For the ac Stark component $H_{\mathrm{ac}}$ we set the laser polarisation at an angle $\beta =45^{\circ}$ with respect to the $z$ axis.}
\label{fig:Hamiltonian}
\end{figure*}

We now briefly describe the process that the \texttt{build\_hamiltonians} function follows. The first step of the calculation is to populate nine two-dimensional \texttt{numpy.ndarray} objects using the \texttt{generate\_vecs} function, one for each Cartesian component of the three angular momentum vectors. To perform this population we use the standard definition of the ladder operators and the relations between spherical and Cartesian components of a general angular momentum $j$:

\begin{subequations}

\begin{gather}
\bra{j',m_{j}'}j_{+}\ket{j,m_{j}}=\sqrt{j(j+1)-m_{j}m'_{j}}
\delta _{j,j'}\delta _{m_{j}',m_{j}+1},
\\
j_{-} = j_{+}^{\dagger},
\\
j_{x}=\frac{1}{2}\left (j_{+} + j_{-}\right ),
\\
j_{y}=\frac{i}{2}\left ( j_{-} - j_{+}\right ),
\\
j_{z}=\frac{1}{2}\left ( j_{+}\cdot j_{-} - j_{-}\cdot j_{+}\right ).
\end{gather}
\end{subequations}
Following this initialisation step, we store each component of the various $j$ as elements in a length-3 list. The final initialisation step is to create a single unified state space, which is done via repeated use of \texttt{numpy.kron} and identity matrices of appropriate dimensions. This transforms each of the 3 vectors from their own, independent, state spaces into the combined space of $i_{\mathrm{A}}$, $m_{\mathrm{A}}$, $i_{\mathrm{B}}$, $m_{\mathrm{B}}$.

The remainder of the functions in \texttt{hamiltonian} implement the expressions from section~\ref{sec:theory} with each of the terms in \eqref{eq:total_ham} calculated by a separate function call and the component of $H_{\mathrm{AB}}$ represented by a \texttt{numpy.ndarray} object. These component objects are shown in Fig.~\ref{fig:Hamiltonian} for an exemplar molecule where $i_{\mathrm{A}}=i_{\mathrm{B}}=3/2$ for the first 3 rotational states. The advantage of this approach is that, having abstracted the angular momentum to a vector once, it need not be repeated; this allows a speed-up using the fast vector processing of numpy and scipy. Similarly, for the terms in $H_{\mathrm{ext}}$, we initially treat the electric, magnetic or optical fields as unit vectors, such that these terms can be scaled later.

To assemble the total Hamiltonian, $H_{\mathrm{rot}}$ and $H_{\mathrm{hf}}$ are simply added together to form $H$. To calculate $H_{\mathrm{ext}}$, we first construct field-independent matrices $H_{\mathrm{Z}}/|{\boldsymbol{B}}|$, $H_{\mathrm{dc}}/|\boldsymbol{E}_\mathrm{dc}|$, and $H_{\mathrm{ac}}/I_\mathrm{ac}$. These are then combined to form $H_{\mathrm{ext}}$, by multiplying them by $|\boldsymbol{B}|$, $|\boldsymbol{E}_\mathrm{dc}|$, and $I_\mathrm{ac}$ as required. Each of these three terms is then added to $H$ to form the total Hamiltonian $H_{\mathrm{AB}}$. To diagonalise the Hamiltonian, we recommend using \texttt{numpy.linalg.eigh} as it can not only handle simultaneous multi-processing of multiple field magnitudes but also exploits the Hermitian property of the Hamiltonian matrix to speed up calculations. This function is based on the \texttt{\_syevd} and \texttt{\_heevd} routines in the LAPACK linear algebra package for Fortran 90. As an example, to generate and diagonalize the Hamiltonians needed for a Zeeman plot covering a magnetic field range of 1 to 500 G, with a constant electric field $E_\mathrm{dc}= 5$~kV\,cm$^{-1}$ and off-resonant light intensity $I_\mathrm{ac}=2.5$ kW\,cm$^{-2}$, we run the script:
\begin{lstlisting}
import numpy
import diatomic.hamiltonian as hamiltonian
from diatomic.constants import Rb87Cs133

Nmax=6
H0,Hz,Hdc,Hac = hamiltonian.build_hamiltonians(Nmax,Rb87Cs133,\
                               zeeman=True,Edc=True,ac=True)

I = 2.5e7 #W / m^2
E = 5e5 #V / m
B = numpy.linspace(1, 500, int(50))*1e-4 #T

H = H0[..., None]+\
    Hz[..., None]*B+\
    Hdc[..., None]*E+\
    Hac[..., None]*I
H = H.transpose(2,0,1)

energies, states = numpy.linalg.eigh(H)
\end{lstlisting} 
The variable \texttt{Rb87Cs133} contains all of the molecular constants for $^{87}$Rb$^{133}$Cs needed for the construction of the Hamiltonian in a Python dictionary:

\begin{lstlisting}
Rb87Cs133 = {"I1":1.5, # nuclear spin of nucleus A
            "I2":3.5, # nuclear spin of nucleus B
            "d0":1.225*DebyeSI, # molecule frame dipole moment
            "Brot":490.173994326310e6*h, # rotational constant
            "Drot":207.3*h, # centrifugal distortion coefficient
            "Q1":-809.29e3*h, # electric quadrupole coupling for A
            "Q2":59.98e3*h, # electric quadrupole coupling for B
            "C1":98.4*h, # Nuclear spin-rotation coefficient for A
            "C2":194.2*h, # Nuclear spin-rotation coefficient for B
            "C3":192.4*h, # Tensor nuclear spin-spin coefficient
            "C4":19.0189557e3*h, # Scalar nuclear spin-spin coeff.
            "MuN":0.0062*muN, # magnetic moment from rotation ang mom
            "Mu1":1.8295*muN, # magnetic moment for A incl. shielding
            "Mu2":0.7331*muN, # magnetic moment for B incl. shielding
            "a0":2020*4*pi*eps0*bohr**3, # isotropic pol, 1064nm
            "a2":1997*4*pi*eps0*bohr**3, # anisotropic pol, 1064nm
            "Beta":0} # laser polarisation angle wrt z
\end{lstlisting} 
All quantities in the dictionary are defined in SI units. We include sample dictionaries in an additional module $\texttt{constants.py}$ from which we import \texttt{Rb87Cs133} in this example. To perform the equivalent calculation for other molecules either a different set of constants can be imported from $\texttt{constants.py}$ or a custom dictionary should be defined. Note that \texttt{diatomic-py} cannot calculate the values of $\alpha ^{(0)}$ and $\alpha ^{(2)}$ and so these must be supplied, by the user, for each wavelength $\lambda $. Where available the constants we have supplied are for $\lambda = 1064~\text{nm}$.

\subsection{Calculate}
\label{sec3.2}

\texttt{Calculate} contains functions that deal with the result of\break the Hamiltonian diagonalisation. This includes a set of three functions \texttt{label\_states\_N\_MN}, \texttt{label\_states\_I\_MI}, \texttt{label\_states\_F\_MF}. These take the array of eigenvectors and evaluate the expectation values of $\boldsymbol{N}\cdot\boldsymbol{N}$ and $N_z$; $\boldsymbol{I}\cdot\boldsymbol{I}$ and $I_z$; and $\boldsymbol{F}\cdot\boldsymbol{F}$ and $F_z$, respectively, and can be used to assign quantum numbers to the eigenstates. Here $\boldsymbol{I}=\boldsymbol{I}_\mathrm{A}+\boldsymbol{I}_\mathrm{B}$ is the operator for the total nuclear spin, and $\boldsymbol{F}=\boldsymbol{N}+\boldsymbol{I}$ is that for the total angular momentum.  

The function \texttt{transition\_dipole\_moment} calculates the transition dipole moment (in units of the molecule-frame dipole moment) between one eigenstate $\ket{i}$ and a range of others $\ket{f}$. This calculation is performed by first constructing the space-fixed electric dipole operator $\boldsymbol{\mu}$ and then calculating $\bra{i}\boldsymbol{\mu}\ket{f}$ by matrix multiplication. We also include the functions \texttt{magnetic\_moment} and \texttt{electric\_moment} that calculate the lab-frame magnetic and electric dipole moments (in SI units) for each eigenstate. Each of these functions constructs the appropriate dipole moment operator $\boldsymbol{\mu} _{z}$ for either the electric or magnetic dipole moment pointed along the quantisation axis ($z$) before calculating the expectation value $\bra{i}\boldsymbol{\mu} _{z}\ket{i}$.

Finally, we include a function for alternative ordering of the energy levels. By default the energy levels are returned in order of ascending energy, such that two levels that cross one another exchange indices. To prevent this, and return levels with indices that reflect the character of the states rather than energy, we provide the function \texttt{sort\_smooth}. This orders energy levels such that, for each state $\psi ^{p}$, the overlap $\braket{\psi ^{p}_{k}|\psi ^{p}_{k+1}}$ is maximal, where $k$ is an index that increments with the independent variable of the calculation. For each $k$ the function calculates the overlap of each eigenstate $\psi ^{p}_{k}$ with all others for $k+1$ i.e. the matrix product $C_{k}^{\mathrm{T}} C_{k+1} = O_{k}$ where $C_{k}$ is the eigenvector matrix at field $k$. By finding the index $q$ for which the value of $\bra{q}O_{k}\ket{p} $ is maximal, we can locate which pair of eigenstates has the largest overlap. If $p\neq q$ then we infer that two energy levels have crossed and swap the indices such that $\psi ^{q}_{k+1} \leftrightarrow \psi ^{p}_{k+1}$ and similarly with the eigenenergies $E^{q}_{k+1} \leftrightarrow E^{q}_{k+1}$.

\subsection{Benchmarking}
\label{sec3.3}

\begin{figure*}[t]
\includegraphics[width=\textwidth]{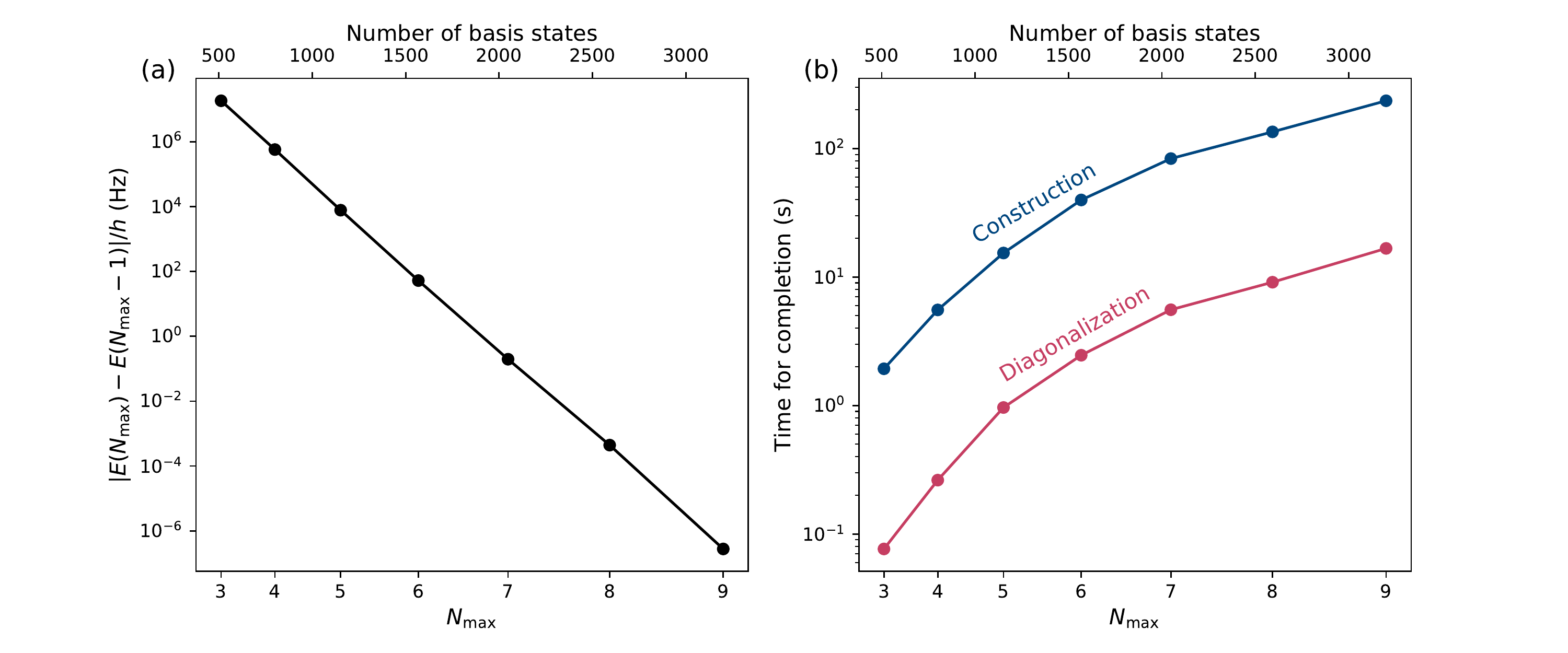}
\caption{Convergence and code run time. We perform a calculation for $^{87}$Rb$^{133}$Cs in a magnetic field $B=181.5$~G and an electric field $E_\mathrm{dc}=5$~kV\,cm$^{-1}$, with an optical field of intensity $I_\mathrm{ac}=2.5$~kW\,cm$^{-2}$ and with $\lambda =1064$~nm, that is linearly polarised orthogonal to the magnetic and electric fields. (a)~Convergence of the calculated energy for the spin-stretched lowest-energy state $(N=0, M_{F}=5)$ as a function of the number of basis states included in the calculation. (b)~Time taken by each calculation; each marker is the average time across 5 calculations performed using an Intel i5-8350U CPU @ 1.7 GHz with 8 GB of RAM.}
\label{fig:timing}
\end{figure*}

\begin{figure*}[t]
\includegraphics[width=\textwidth]{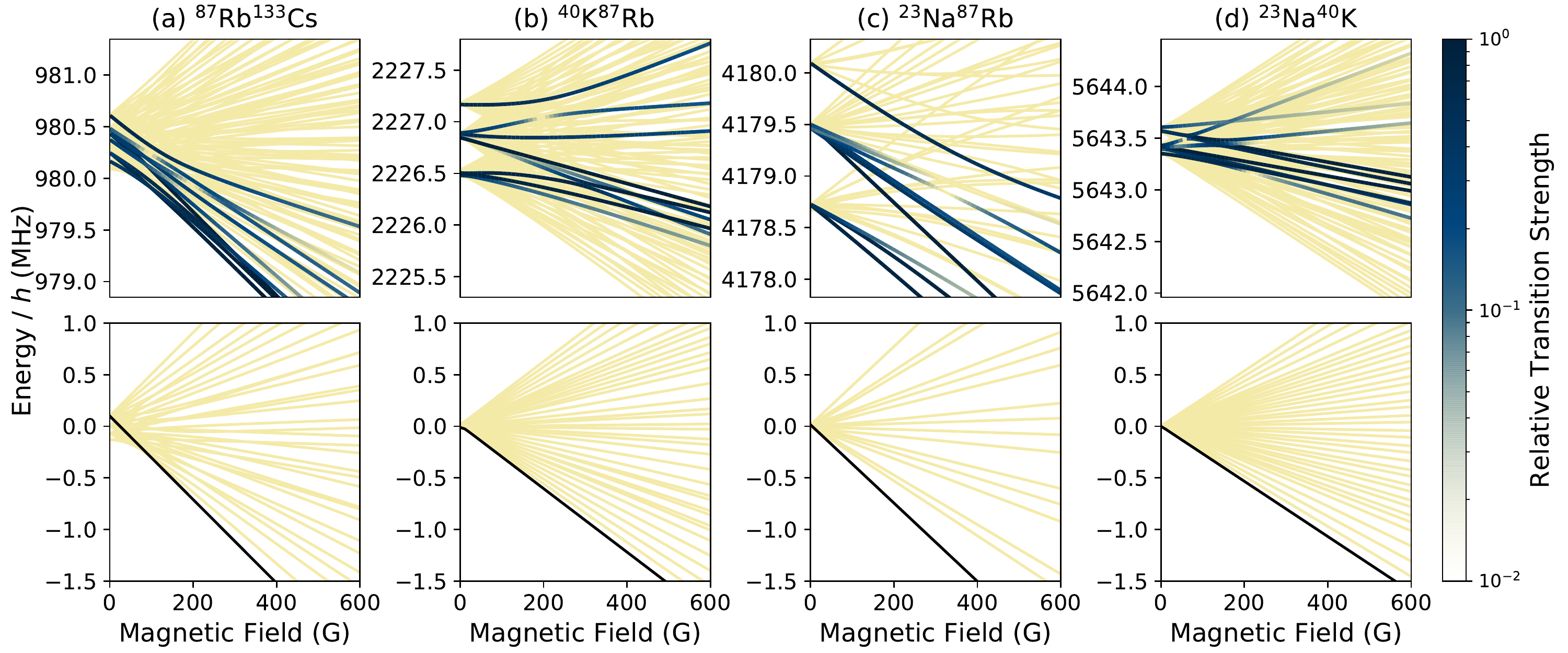}
\caption{Zeeman structure of states with $N=0$ and 1 for a selection of bialkali molecules: (a)~$^{87}$Rb$^{133}$Cs, (b)~$^{40}$K$^{87}$Rb, (c)~$^{23}$Na$^{87}$Rb, (d)~$^{23}$Na$^{40}$K. The hyperfine structure for $N=0$ is shown in the lower panels, with the high-field hyperfine ground state indicated in black. The structure for $N=1$ is shown in the upper panels. The relative transition strengths for one-photon transitions from the high-field hyperfine ground state are shown by the blue colour map.}
\label{fig:Zeeman}
\end{figure*}

The accuracy of the calculation increases with the number of basis states included in the calculation. However, this also increases the time for computation. The dc Stark component of the Hamiltonian causes the largest mixing between rotational states, and therefore calculations for molecules in large dc electric fields are the most sensitive to the number of states included in the calculation. For the purpose of these tests we use the molecule \textsuperscript{87}Rb\textsuperscript{133}Cs; we do not anticipate that there will be much variation in the convergence or time for different bialkali species.

We control the number of basis states by changing the variable $N_{\mathrm{max}}$, i.e. the quantum number of the highest-energy rotational state included in the calculation. For each result, we plot the change in energy of the state when the new rotational state is included in the calculation.

For our example, we consider the $^{87}$Rb$^{133}$Cs molecule in a magnetic field $B=181.5~\mathrm{G}$, a dc electric field $E_\mathrm{dc}=5$~kV\,cm$^{-1}$, and an optical field, with $\lambda =1064$ nm, with laser intensity ${I_\mathrm{ac}=2.5~\mathrm{kW}\,\mathrm{cm}^{-2}}$, that is linearly polarised orthogonal to the magnetic and electric fields. This is a typical configuration at an energy scale relevant to prior experimental work~\cite{Blackmore2020}. We find that for this configuration, the energy of $N=0$ changes by less than $h\times1$~Hz for $N_{\mathrm{max}}>7$, better than the uncertainty on experimental measurements of the absolute binding energy~\cite{Molony2016} or typical uncertainty in rotational spectroscopy~\cite{Gregory2016,Blackmore2020a}.

In Fig.~\ref{fig:timing}(b), we show the time that each calculation took using an Intel i5-8350U CPU @ 1.7 GHz with 8 GB of RAM. We break the run time up into the construction phase, where the Hamiltonian is constructed, and the numerical diagonalization using \texttt{numpy.linalg.eigh}. We see that the Hamiltonian construction takes an order of magnitude longer than the diagonalization for all calculations. However, for a calculation with multiple field magnitudes, this construction step must be performed only once. We anticipate that the duration of the diagonalization stage would scale linearly with the number of field magnitudes being studied and as the cube of the number of basis states i.e. as $(N_{\mathrm{max}}+1)^{6}$. Note that the largest calculations shown, with $N_{\mathrm{max}} =9$, take only a few minutes in total to complete, demonstrating the utility of the code without access to high-performance computing facilities.

\section{Examples}
\label{sec:examples}
In this section we briefly describe example calculations that are relevant to current research into controlling the quantum states of ultracold bialkali molecules.
\begin{figure*}[t]
	\includegraphics[width=\textwidth]{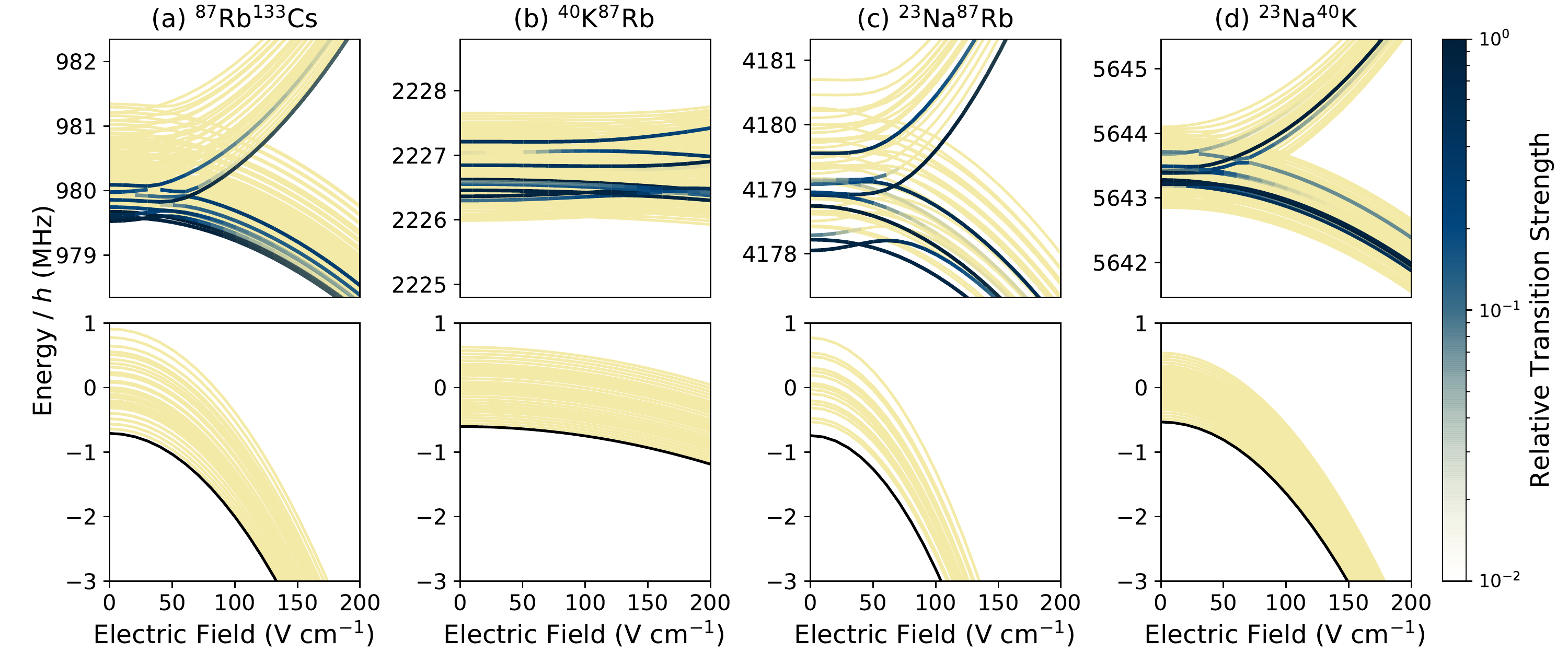}
	\caption{dc Stark shift for the $N=0,1$ for a selection of bialkali molecules: (a)~$^{87}$Rb$^{133}$Cs, (b)~$^{40}$K$^{87}$Rb, (c)~$^{23}$Na$^{87}$Rb, (d)~$^{23}$Na$^{40}$K. The electric field is applied in addition to a parallel dc magnetic field of 200 G. The $N=0$ hyperfine structure is shown in the lower panels, with the high-field hyperfine ground state indicated in bold. The $N=1$ structure is shown in the upper panels. The relative transition strength is coded as in Fig.~\ref{fig:Zeeman}.}
	\label{fig:dcStark}
\end{figure*}
\subsection{Zeeman and dc Stark effects}
\label{sec4.1}
In Fig.~\ref{fig:Zeeman} we show the Zeeman structure for $N=0,1$ for a selection of bialkali molecules up to 600 G in increments of 10 G. To perform the calculations, we generate a 3D numpy array constructed by layering the 2D Hamiltonian from each magnetic field to be evaluated, and simultaneously diagonalising all Hamiltonians in a single call of \texttt{numpy.linalg.eigh}. For each molecule, we highlight the spin-stretched hyperfine state that becomes the absolute ground state in the limit of large magnetic field, and also the states with $N=1$ that are connected to it by an allowed one-photon transition. In Fig.~\ref{fig:dcStark}, we show similar dc Stark maps calculated using a similar approach for the same molecules in a 200 G magnetic field. For each calculation, we take the field from $0~\mathrm{V\,cm^{-1}}$ to $250~\mathrm{V\,cm^{-1}}$ in increments of $2.5~\mathrm{V\,cm^{-1}}$.

\subsection{Transition energies and transition dipole moments}
\label{sec:TDM}

\begin{figure}[t]
\centering
\includegraphics[width=0.5\textwidth]{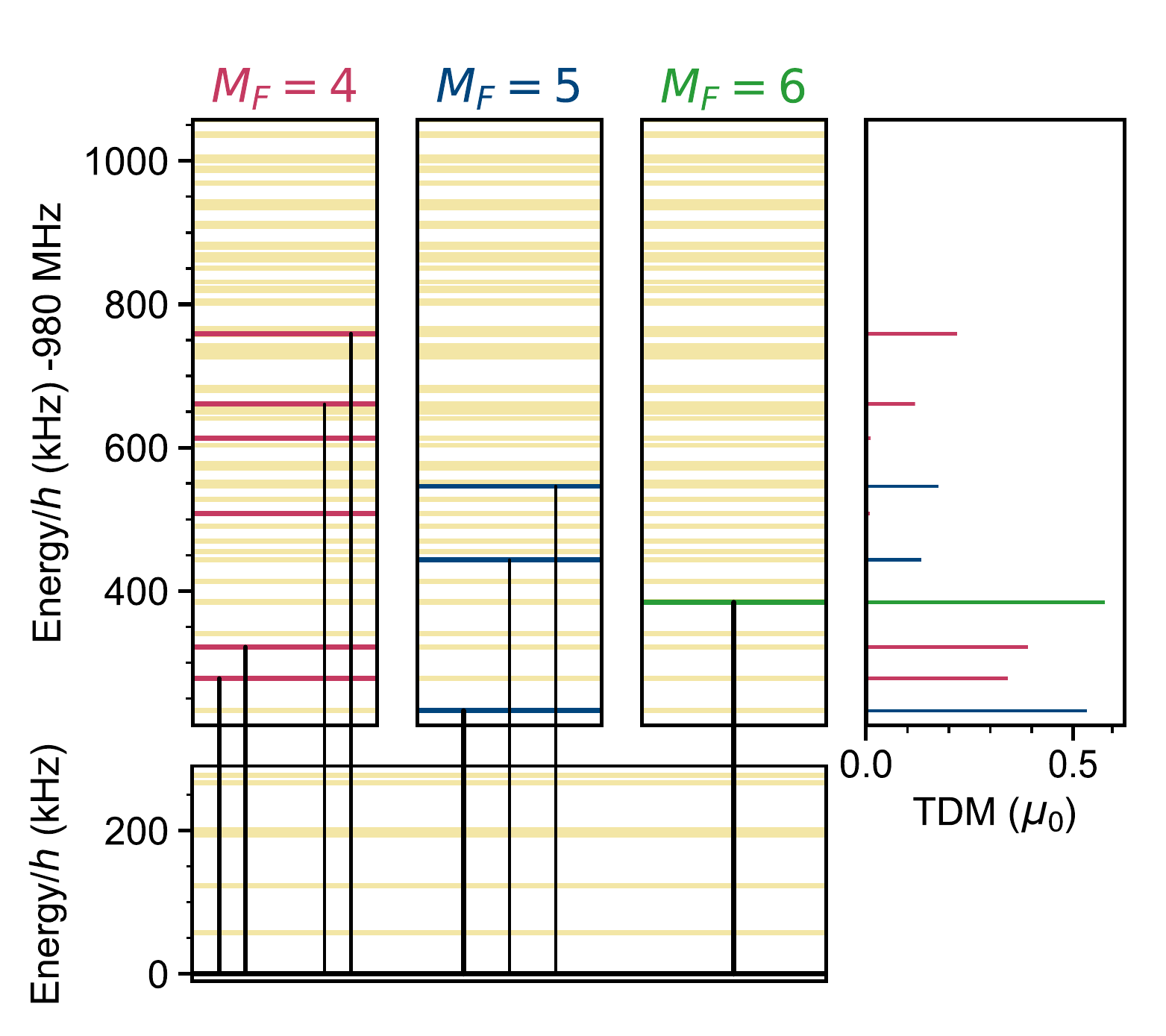}
\caption{Energies for states of $^{87}$Rb$^{133}$Cs with $N=0$ and 1, and allowed transitions from the absolute ground state $(N=0, M_{F}=5)$ at a magnetic field of 181.5 G. States in $N=1$ with allowed one-photon transitions are highlighted and labelled as $M_{F}=4,5,6$. The panel on the right shows the transition dipole moment (TDM) for each transition as a fraction of the molecule-frame dipole moment ($\mu _{0}$).}
\label{fig:transitions}
\end{figure}

Electric dipole transitions between rotational states may be driven using resonant microwaves. In Fig.~\ref{fig:transitions} we show the hyperfine states of $^{87}$Rb$^{133}$Cs in a 181.5 G magnetic field. We choose the absolute ground state $(N=0, M_{F}=5)$ as the initial state from which we want to calculate the transition dipole moments of all allowed transitions. From here, there are allowed one-photon microwave transitions to $N=1$ states with $M_{F}=4,5,6$, depending on the polarisation of the driving field with respect to the quantisation axis, as defined by the magnetic field. For each transition we show the energy of the final state with respect to the initial state along with the transition dipole moment in units of the molecule-frame dipole moment (for $^{87}$Rb$^{133}$Cs, $\mu _{0}=1.23$ D~\cite{Molony2014}).

\subsection{ac Stark effects}
\label{sec4.3}

\begin{figure*}[t]
\includegraphics[width=\textwidth]{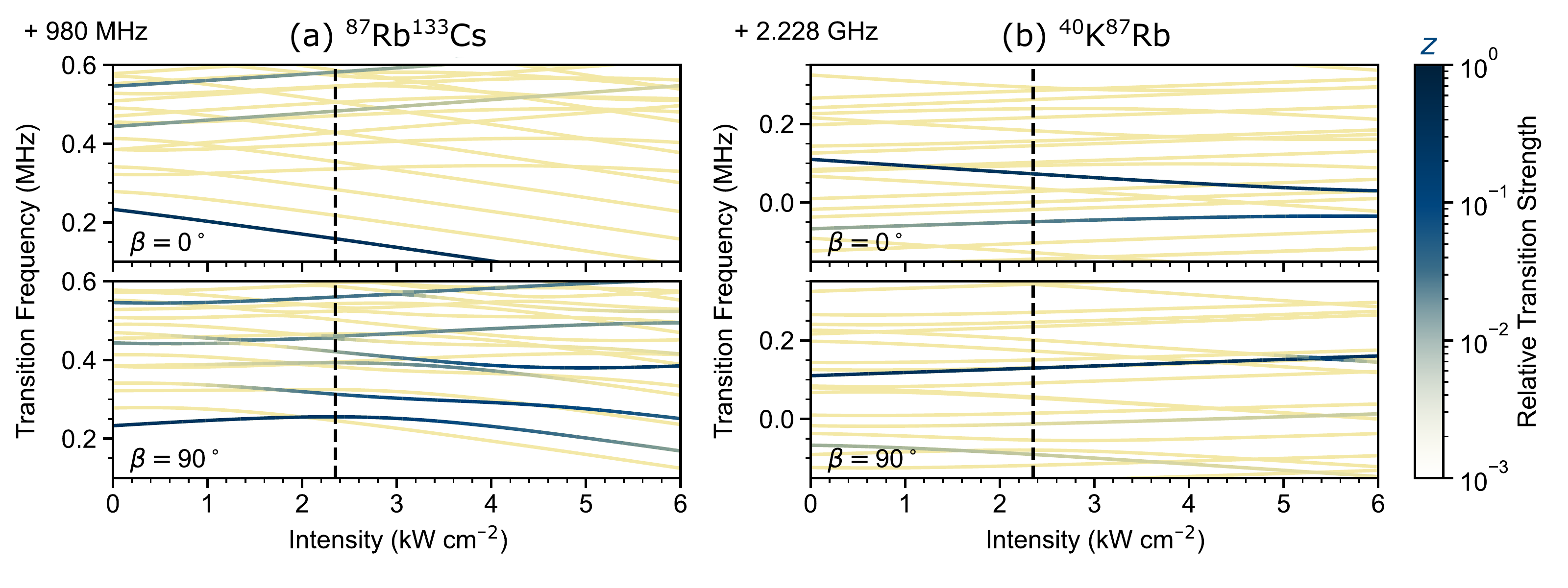}
\caption{The frequency of rotational transitions as a function of off-resonant laser intensity. We show calculations for transitions from (a)~the absolute ground state $\ket{N=0,M_{F}=+5}$ of \textsuperscript{87}Rb\textsuperscript{133}Cs and (b)~the state $\ket{N=0,M_{N}=0,m_{\mathrm{K}}=-4,m_{\mathrm{Rb}}=1/2}$ of \textsuperscript{40}K\textsuperscript{87}Rb. In each case, the off-resonant laser field has a wavelength $\lambda =1064$~nm and is linearly polarised at an angle $\beta $ with respect to $z$. Sub-levels that are not accessible are shown in yellow as a function of laser intensity. The relative transition strengths for microwaves polarised along $z$ are shown as a blue colour map. For the calculations for \textsuperscript{87}Rb\textsuperscript{133}Cs the magnetic field is fixed to be 181.5~G, appropriate for matching the experiments of Blackmore~\textit{et al.}~\cite{Blackmore2020}. In the calculations for \textsuperscript{40}K\textsuperscript{87}Rb the magnetic field is fixed to 549.5~G, appropriate for comparison with the experiments of {Neyenhuis \textit{et al.}~\cite{Neyenhuis2012}}. The dashed black line indicates the intensity used for the calculations shown in Fig.~\ref{fig:AC_Stark_Pi}.} \label{fig:AC_Stark_int}
\end{figure*}

\begin{figure*}[t]
\includegraphics[width=\textwidth]{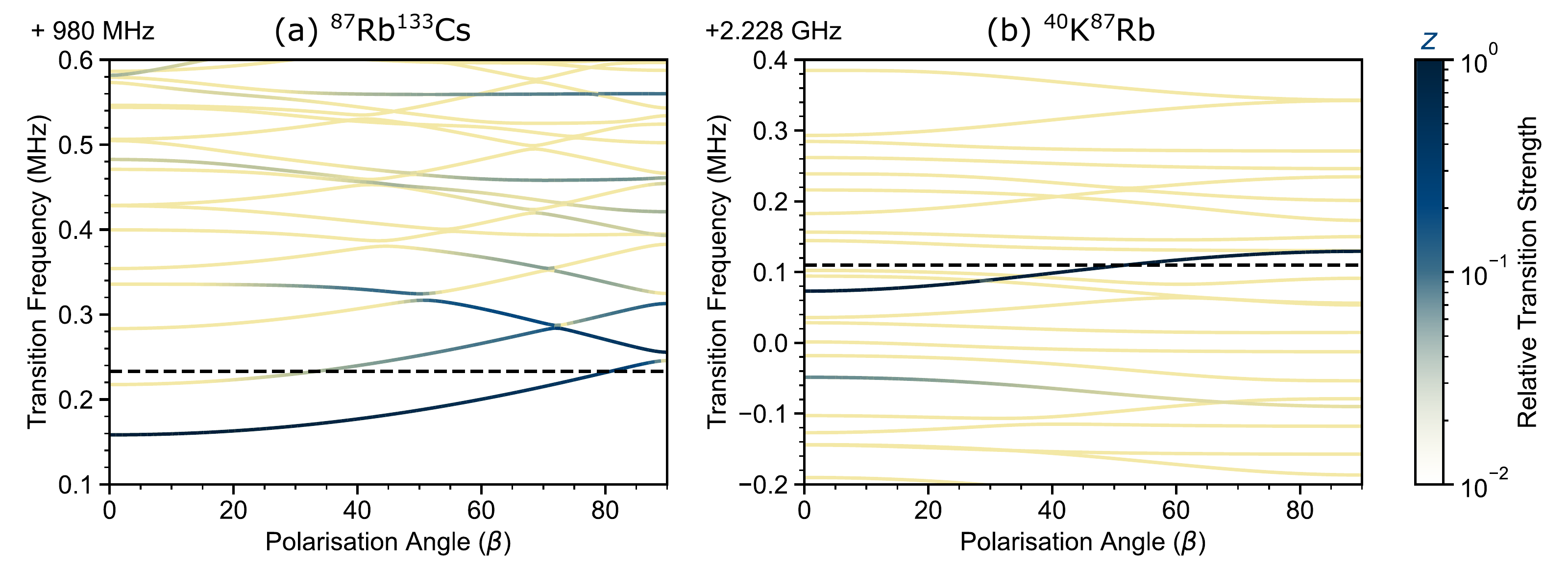}
\caption{The frequency of rotational transitions as a function of off-resonant laser polarisation. We show calculations for transitions from (a) the absolute ground state $\ket{N=0,M_{F}=+5}$ of \textsuperscript{87}Rb\textsuperscript{133}Cs (b) the state $\ket{N=0,M_{N}=0,m_{\mathrm{K}}=-4,m_{\mathrm{Rb}}=1/2}$ of \textsuperscript{40}K\textsuperscript{87}Rb. In each case the off-resonant laser field has a wavelength $\lambda =1064$ nm, an intensity of 2.35 kW\,cm$^{-2}$, and is linearly polarised at an angle $\beta $ with respect to $z$. The relative transition strengths for microwaves polarised along $z$ are shown as a blue colour map. For the calculations shown in (a) the magnetic field is fixed to be 181.5~G, appropriate for matching the experiments of Blackmore~\textit{et al.}~\cite{Blackmore2020}. For the calculations shown in (b) the magnetic field is fixed to 549.5~G, appropriate for comparison with the experiments of Neyenhuis \textit{et al.}~\cite{Neyenhuis2012}. The dashed black lines indicate the transition frequencies for the strongest transition in $N=1$ in the absence of the laser field.}
\label{fig:AC_Stark_Pi}
\end{figure*}

The ac Stark effect is important for optically trapped molecules. Using our code, transition frequencies can be calculated as a function of either laser intensity $I_\mathrm{ac}$ or polarisation angle $\beta $. We give examples of each of these calculations for the molecules $^{87}$Rb$^{133}$Cs and $^{40}$K$^{87}$Rb at experimentally relevant magnetic fields of 181.5~G and 545.9~G respectively, and laser wavelength $\lambda =1064$~nm. In each case we highlight allowed transitions for microwaves polarised along $z$ from an initial state with $N=0$; we choose ${(N=0, M_{F}=5)}$ for $^{87}$Rb$^{133}$Cs and $(N=0, m_{\mathrm{K}}=-4, m_{\mathrm{Rb}}=1/2)$ for $^{40}$K$^{87}$Rb. In Fig.~\ref{fig:AC_Stark_int} we show the relevant transition frequencies as a function of the laser intensity for $\beta =0^{\circ}$ and $90^{\circ}$. For the case of $^{87}$Rb$^{133}$Cs at $\beta =90^{\circ}$ we observe a complex pattern of avoided crossings as the trapping light mixes states with different $M_{F}$. This behaviour matches that observed in experiments~\cite{Gregory2017,Blackmore2020}. In Fig.~\ref{fig:AC_Stark_Pi} we show a similar plot, but varying $\beta $ with fixed ${I_\mathrm{ac}=2.35~\mathrm{kW}\,\mathrm{cm}^{-2}}$. To perform this calculation we recalculate the component $H_{\mathrm{ac}}^{(2)}$ of the Hamiltonian from scratch for each value of $\beta $. The horizontal dashed line shows the transition frequency for the strongest allowed transition in the absence of the trap light. For \textsuperscript{40}K\textsuperscript{87}Rb, there is one strong transition that is allowed across all polarisation angles. This transition frequency crosses the free-space value when $\beta =52^{\circ}$, in agreement with experimental observations of Neyenhuis~\textit{et al.}~\cite{Neyenhuis2012}.

\section{Installation and usage}
\label{sec:Install}

Installation of the package is performed by using the \texttt{.whl} files from the GitHub repository~\cite{github} or directly from the Python Package Index using \texttt{pip install diatomic}.

\section{Conclusions and outlook}
\label{sec:conclusions}

We have presented a Python code that automates the construction of a Hamiltonian that describes the rotational and hyperfine structure of $^{1}\Sigma $ molecules. The Hamiltonian includes terms to describe interactions between the molecule and external dc magnetic, dc electric and the off-resonant optical fields necessary for trapping the molecules. This facilitates the straightforward calculation of Zeeman, dc Stark, and ac Stark maps of the hyperfine structure that can be readily compared with measurements from current experiments. Additional functions for the calculation of static magnetic and electric dipole moments of states are also provided.

Useful future additions to the code may include: greater flexibility in the geometry of the applied fields, e.g. non-parallel dc magnetic and dc electric fields; simulation of dressing by near-resonant microwave fields; extension to $^{2}\Sigma $ molecules, relevant to experiments on laser-cooled molecules.

\section*{Conflicts of Interest}
The authors declare that they have no known competing financial interests or personal relationships that could have appeared to influence the work reported in this paper.

\section*{Acknowledgements}
The authors thank Jes\'{u}s Aldegunde for assistance in writing the code and for stimulating discussions on the mathematics. We also thank Rahul Sawant, Andrew Innes, Luke Fernley and Albert Tao for testing and bug-fixing in early versions of the code.

This work was supported by U.K. {Engineering and Physical Sciences Research Council} (EPSRC) Grants {EP/P01058X/1} and {EP/P008275/1}. JAB acknowledges support from the EPSRC Quantum Computing and Simulation Hub EP/T001062/1.

\bibliographystyle{elsarticle-num}
\bibliography{hyperfine}

\end{document}